\documentclass[apj,numberedappendix]{emulateapj}

\newcommand{\be}{\begin{eqnarray}}
\newcommand{\ee}{\end{eqnarray}}
\newcommand{\lp}{\left(}
\newcommand{\rp}{\right)}

\newcommand{\E}[1]{\times10^{#1}}
\newcommand{\smpy}{ \ M_\odot \ {\rm yr}^{-1}}
\newcommand{\cgsp}{ \ {\rm dyne \ cm}^{-2} }
\newcommand{\cgsd}{ \ {\rm g \ cm}^{-3} }

\shorttitle{HELIUM BURNING ON WDs}
\shortauthors{SHEN \& BILDSTEN}

\begin{document}


\title{Unstable helium shell burning on accreting white dwarfs}

\author{Ken J. Shen}
\affil{Department of Physics, Broida Hall, University of California, Santa Barbara, CA 93106; kenshen@physics.ucsb.edu}

\author{Lars Bildsten}
\affil{Kavli Institute for Theoretical Physics and Department of Physics, Kohn Hall, University of California, Santa Barbara, CA 93106; bildsten@kitp.ucsb.edu}


\begin{abstract}

AM Canum Venaticorum (AM CVn) binaries consist of a degenerate helium donor and a helium, C/O, or O/Ne WD accretor, with accretion rates of $\dot{M} = 10^{-13} - 10^{-5} \smpy$.  For accretion rates $<10^{-6} \smpy$, the accreted helium ignites unstably, resulting in a helium flash.  As the donor mass and $\dot{M}$ decrease, the ignition mass increases and eventually becomes larger than the donor mass, yielding a ``last-flash'' ignition mass of $\lesssim 0.1 \ M_\odot$.  \cite{bswn07} predicted that the largest outbursts of these systems will lead to dynamical burning and thermonuclear supernovae.  In this paper, we study the evolution of the He-burning shells in more detail.  We calculate maximum achievable temperatures as well as the minimum envelope masses that achieve dynamical burning conditions, finding that AM CVn systems with accretors $\gtrsim 0.8 \ M_\odot$ will undergo dynamical burning.  Triple-$\alpha$ reactions during the hydrostatic evolution set a lower limit to the $^{12}$C mass fraction of $0.001 - 0.05$ when dynamical burning occurs, but core dredge-up may yield $^{12}$C, $^{16}$O, and/or $^{20}$Ne mass fractions of $\sim 0.1$.  Accreted $^{14}$N will likely remain $^{14}$N during the accretion and convective phases, but regardless of $^{14}$N's fate, the neutron-to-proton ratio at the beginning of convection is fixed until the onset of dynamical burning.  During explosive burning, the $^{14}$N will undergo $^{14}$N$(\alpha,\gamma)^{18}$F$(\alpha,p)^{21}$Ne, liberating a proton for the subsequent $^{12}$C$(p,\gamma)^{13}$N$(\alpha,p)^{16}$O reaction, which bypasses the relatively slow $\alpha$-capture onto $^{12}$C.  Future hydrodynamic simulations must include these isotopes, as the additional reactions will reduce the Zel'dovich-von Neumann-D\"{o}ring (ZND) length, making the propagation of the detonation wave more likely.

\end{abstract}



\section{Introduction}

Helium accretes onto helium, C/O, and O/Ne white dwarfs (WDs) at accretion rates ranging from $\dot{M} = 10^{-13} - 10^{-5} \smpy$ in AM Canum Venaticorum (AM CVn) binaries.  \cite{bswn07} showed that the resulting unstable helium flashes for $\dot{M}$'s between $ 10^{-8} - 10^{-6} \smpy$ may lead to a new kind of thermonuclear supernova (SN) that rises and falls in $\simeq 10 $ days and reaches at most one-tenth the brightness of a normal Type Ia SN.  These .Ia SNe have yet to be detected (the recent SN 2008ha comes the closest to resembling a .Ia SN; \citealt{fol09}), but upcoming surveys such as the Palomar Transient Factory, Pan-STARRS-1, and Sky-Mapper will have cadences, depths, and sky coverage adequate to detect them if \cite{bswn07}'s predicted rates in old stellar populations of $\sim 1$ per $10^4$ yr in a $10^{11} \ M_\odot$ galaxy are correct.

While it is known that the accumulated helium layer on the WD burns in an unstable manner once $\dot{M} \lesssim 10^{-6} \smpy$ \citep{it89}, it remains uncertain as to how the resulting flash evolves.  In this paper, we calculate the time evolution of fully convective He-burning envelopes as a series of hydrostatic solutions.  This convective burning phase lasts for months $-$ decades, much longer than the time it takes for transverse equilibration.  Hence, we assume spherical symmetry throughout our work.  Due to hydrostatic radial expansion, the pressure at the base of the accumulated shell decreases, which leads to a maximum temperature for the base of the He-burning shell.  For each WD mass, there is a minimum helium shell mass for which the nuclear heating timescale becomes shorter than the time for a sound wave to traverse the radial thickness of the shell.  Though this is not a definitive condition for the onset of a rapid detonation, it certainly is the beginning of dynamical burning, either via a deflagration or a detonation.  These minimum shell masses are naturally reached in the evolution of the AM CVn binaries \citep{bswn07}.  Envelopes smaller than these do not result in hydrodynamic burning as we describe above but instead complete their burning in a manner that allows for an accurate calculation with a hydrostatic code.  Though mass loss occurs in these helium novae, like that which likely caused the outburst of V445 Puppis \citep{kh03,in08,kato08}, such an event is not dynamically explosive.

Our work provides the initial data relevant to the hydrodynamic calculations needed to determine the outcome of these unstable flashes.  In \S \ref{sec:mign}, we describe previous calculations of helium ignition masses that motivate our range of envelope masses.  In \S \ref{sec:calc}, we describe our hydrostatic calculation and calculate the maximum temperatures reached during the envelope's hydrostatic evolution, as well as the minimum envelope masses required to achieve dynamical burning.  For AM CVn binaries with accretor masses $ \gtrsim 0.8 \ M_\odot$, the last (and largest) helium flash will undergo dynamical burning.  While the focus of this paper is on the evolution of He-burning shells in AM CVn binaries, our method's simplicity also lends itself to other applications, such as the helium core flash at the tip of the red giant branch (RGB), which we explore in \S \ref{sec:RGB}.  The effect of composition on our results is described in \S \ref{sec:comp}.  We find that conditions for $\alpha$-chain and $^{14}$N burning are not reached prior to dynamical burning, leaving unburned $^{12}$C, $^{14}$N, $^{16}$O, and $^{20}$Ne that will make a propagating detonation more likely in the thin shell as compared to calculations assuming pure helium.  The nuclear burning will be even further enhanced by the catalyzing effect of protons released in $^{18}$F $\alpha$-captures.  We also find that the neutron-to-proton ratio is fixed between the beginning of convection and the onset of dynamical burning.  In \S \ref{sec:conc}, we summarize our work.


\section{Helium ignition masses on accreting white dwarfs}
\label{sec:mign}

Though the purpose of this paper is to study the evolution of unstable helium flashes on helium, C/O, and O/Ne WDs, we must first set the stage by discussing the expected ignition masses, $M_{\rm ign}$.  All work done to date has been for constant $\dot{M}$'s and a range of core temperatures, $T_c$, which are not necessarily the appropriate conditions for AM CVn binaries.  However, these initial works prove adequate to reveal the parameters relevant for the AM CVn's.


\subsection{Numeric work and analytic expectation}
\label{sec:num_mign}

There have been many numerical calculations of the ignition masses of He-accreting WDs \citep{ns77,taam80a,taam80b,nom82a,fs82,wtw86,it91,lt91,ww94,pier99,pct01,gor02,yl04}, covering a large range of WD core masses ($ M_c = 0.4 - 1.28 \ M_\odot$) and accretion rates ($\dot{M} =  5\E{-10} - 2\E{-7} \smpy$).  Note that the conductive opacity used in all of the pre-1991 studies was erroneously high due to a typographical error in \cite{iben75}.  As \cite{it91} note, a higher conductive opacity reduces the heat flow into the degenerate core, so the envelope stays hotter than it should be, and thus the pre-1991 ignition masses are as much as a factor of 2 lower than they should be.

Some authors only publish the accreted mass at ignition, $M_{\rm ign}$, which can be larger than the mass initially contained in the convective zone due to the fact that the temperature peak in the hot accreting envelope can be above the envelope base.  Thus, for $\dot{M} > 3\E{-8} \smpy$, the convective envelope can be initially as small as $0.5 \ M_{\rm ign}$ \citep{fs82,lt91}, and so $M_{\rm ign}$ is an upper limit to the convective envelope mass, $M_{\rm env}$.  For $\dot{M} \lesssim 10^{-8} \smpy$, the envelope is not drastically hotter than the core, so the temperature peak at ignition is near the envelope base, and $M_{\rm env} \simeq M_{\rm ign}$ \citep{lt91,it91}.

\begin{figure}
	\plotone{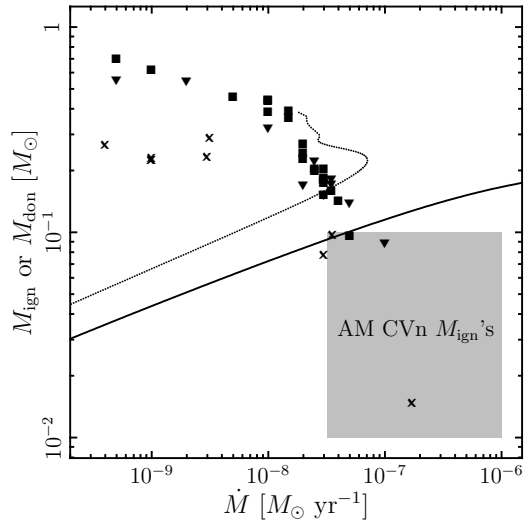}
	\caption{The points show $M_{\rm ign}$ as a function of $\dot{M}$ for $0.6 \ M_\odot$ (\emph{squares}), $0.8 \ M_\odot$ (\emph{triangles}), and $1.08-1.2 \ M_\odot$ WDs (\emph{crosses}) from numerical studies in the published literature; see text for sources.  The values for the $1.08-1.2 \ M_\odot$ models may be too low due to an opacity error (see text).  The donor mass evolution of a degenerate helium donor from C. Deloye (2007, private communication) is shown as a solid line; the donor mass evolves from upper-right to lower-left.  Ignition masses expected for AM CVn binaries are within the shaded region.  The evolution of a He-burning donor is shown as a dotted line, evolving first to higher and then to lower $\dot{M}$ \citep{yung08}.}
	\label{fig:num_mign}
\end{figure}

Figure \ref{fig:num_mign} displays numerically calculated ignition masses as a function of $\dot{M}$, presumed to be constant, for $0.6 \ M_\odot$ (\emph{squares}), $0.8 \ M_\odot$ (\emph{triangles}), and $1.08-1.2 \ M_\odot$ (\emph{crosses}) WDs.  The $1.08-1.2 \ M_\odot$ ignition masses \citep{taam80b,nom82a,fs82,wtw86} were calculated before 1991, so they may be lower than an updated calculation would find.  Note that the numerical ignition masses are nearly independent of $M_c$ for $0.6 \ M_\odot \leq M_c \leq 1.2 \ M_\odot$ and $\dot{M}>10^{-8} \smpy$.  The data points shown in Figure \ref{fig:num_mign} do not include the NCO chain \citep{hashi86,pier99,pct01}, which can ignite a flash prior to the onset of triple-$\alpha$ burning but is likely negligible for AM CVn systems; see \S \ref{sec:comp_acc} for details.

\cite{it89} used a one-zone calculation to estimate the helium ignition mass as
\be
	M_{\rm ign} &\simeq& 0.04 \ M_\odot \lp \frac{R_b}{5\E{8} \ {\rm cm} } \rp^{3.75}  \nonumber \\
	&& \times \lp \frac{M_c}{M_\odot} \rp^{-0.30} \lp \frac{\dot{M}}{10^{-8} \smpy} \rp^{-0.57} ,
\ee
where $R_b$ is the radius at the base of the accreted envelope.  This rough calculation underestimates the ignition masses from the numerical work listed above by a factor of $\sim 5$, or even a factor of 10 for $M_c > M_\odot$.  However, the scaling of $M_{\rm ign} \propto \dot{M}^{-0.57}$ roughly matches the numerical results for $\dot{M} > 10^{-8} \smpy$.

In addition to their numerical work, \cite{it91} attempted to improve their previous one-zone analytic estimate by accounting for the inward luminosity into the core from the envelope.  Their result is
\be
	M_{\rm ign} &\simeq& 0.07 \ M_\odot \lp \frac{\beta}{0.5} \rp^{-5/7} \lp \frac{R_b}{5\E{8} \ {\rm cm}} \rp^{8/7} \nonumber \\
	&& \times \lp \frac{M_c}{M_\odot} \rp^{3/7} T_8^{5/7} ,
\ee
where the helium ignition temperature in units of $10^8$ K, $T_8$, is roughly equal to 1, and $\beta$ is the fraction of the gravitational energy released by the accreting envelope material falling that goes into heating the core.  Note that this $M_{\rm ign}$ is independent of $\dot{M}$, which is in conflict with the numeric work listed above for $\dot{M}>3\E{-9} \smpy$.  However, the more gradual scaling with $M_c$ and $R_b$ is in better agreement with the literature than the analytic estimate from \cite{it89}.


\subsection{AM CVn ignition masses}
\label{sec:amcvn_mign}

The evolution of a fully-degenerate helium donor, as appropriate for AM CVn systems, is shown as a solid line in Figure \ref{fig:num_mign} (C. Deloye 2007, private communication), beginning at $\dot{M} > 10^{-6} \smpy $ and evolving to lower accretion rates.  For $\dot{M} \gtrsim 3\E{-6} \smpy $, outside of the bounds of Figure \ref{fig:num_mign}, the He-burning in the accreted shell is thermally stable \citep{it89,ty96,sb07}, potentially resulting in short-period supersoft sources \citep{vt97}.  As the donor loses mass, $\dot{M}$ decreases, the donor evolves from the upper-right of Figure \ref{fig:num_mign} to the lower-left, and the He-burning becomes unstable \citep{bild06}.  \cite{bswn07} showed that the resulting ignition masses lie within the shaded region.  When the donor mass becomes smaller than the required helium ignition mass, there will be no more helium explosions in the system.  Thus, there is a ``last-flash'' $M_{\rm ign}$ that is the maximum ignition mass for a given $M_c$ in an AM CVn system \citep{bswn07}, which Figure \ref{fig:num_mign} shows to be $\lesssim 0.1 \ M_\odot$.  Accurate $M_{\rm ign}$ values await calculations with $\dot{M}$-evolution during the period of accumulation.  These must also include the core heating from earlier phases of stable burning and weak flashes \citep{ty96,bild06}.

The AM CVn last-flash $M_{\rm ign}$ is lower than the $0.2-0.6 \ M_\odot$ helium shells expected in binary systems with He-burning donors, which were studied as potential Type Ia SN progenitors via the double-detonation scenario \citep{nom82a,nom82b,wtw86,liv90,lg90,lg91,ww94}.  The donor evolution for a He-burning star is shown as a dotted line in Figure \ref{fig:num_mign} \citep{yung08}.  The initial donor and accretor masses are $0.4 $ and $0.6 \ M_\odot$.  The system comes into contact at an orbital period of $80$ min and evolves first to higher $\dot{M}$'s and then to lower $\dot{M}$'s under the influence of gravitational wave radiation.  The $\dot{M}$-evolution for these stars leads to $M_{\rm ign} \gtrsim 0.2 \ M_\odot$, significantly higher than those expected in AM CVn systems.  The strong overlap of the calculated points in Figure \ref{fig:num_mign} with the He-burning star evolution is due to the substantial work on double-detonations.


\section{Calculation and example}
\label{sec:calc}

AM CVn ignition masses can be as large as $0.1 \ M_\odot$, and because the accreted envelope can be such a large proportion of the WD mass, it is important that the envelope and core variables are calculated consistently so as to maintain hydrostatic equilibrium.  Our calculations thus differ from the analytical work of \cite{fs82}, who assumed that the radius of the underlying WD remains unchanged throughout the evolution of the flash.

The lowest density in the underlying WD, just under the base of the accreted envelope, is $\sim 10^5 \cgsd $ at the time of helium ignition.  The associated Fermi energy is equal to the thermal energy when $T = 4 \E{8} {\rm \ K} \ ( \rho / 10^5 \cgsd )^{2/3} $.  Thus, for a core temperature $T_c < 10^8$ K, the temperature will not significantly affect the core structure.  For simplicity, we choose $T_c=10^7$ K for all of our models (unless otherwise noted as in \S \ref{sec:RGB}) as expected for the majority of the helium flashes and a factor of a few lower than the maximum $T_c$ expected for the last flash in AM CVn systems \citep{bild06}.  The core is assumed to have a homogenous composition of 50\% C and 50\% O by mass, but the core structure is nearly independent of the composition choice because $^4$He, $^{12}$C, $^{16}$O, and $^{20}$Ne all have the same charge-to-mass ratio.  The equation of state \citep{scvh95,ts00,rn02}, opacity \citep{ir93,ir96}, nuclear burning network (\citealt{tim99}, unless otherwise noted), neutrino cooling \citep{itoh96}, and electron screening \citep{grab73,aj78,itoh79} are calculated with the MESA code package.\footnote{http://mesa.sourceforge.net/}

The density and radius are integrated outwards in mass coordinates from a central density, $\rho_c$, under the assumption of hydrostatic equilibrium until the desired WD core mass, $M_c$, is reached.  Specifying $\rho_c$, $M_c$, and $T_c$ gives the pressure and radius at the WD core-envelope interface.  Since the pressure and radius are continuous variables, these are equal to the pressure and radius at the base of the accreted envelope, $P_b$ and $R_b$, whereas the temperature and composition are discontinuous at the WD core-envelope boundary by construction.  The assumption of a discontinuous composition relies on minimal diffusion during the accretion phase, justified in Appendix \ref{sec:diff}.

The actively-burning envelope is an efficient convection zone,\footnote{The convective speed, $v_c$, can be defined via the convective flux as $F_c \sim \rho v_c^3$.  Efficient convection is equivalent to the condition that $v_c$ is much smaller than the sound speed.} so that the temperature profile is given by the adiabatic index, $ \nabla_{\rm ad} \equiv \left. \partial \ln T / \partial \ln P \right|_s $.  The pressure and radius at the base of the envelope, calculated from the core integration, are used as the starting point of the envelope integration.  A base temperature, $T_b$, is specified, and the envelope variables are integrated outwards until $P(r) \ll P_b $, yielding the envelope mass, $M_{\rm env}$, as a function of $T_b$ for a particular combination of $P_b$ and $R_b$.  For a given $M_c$, there is at most one $T_b$ for each $P_b$ that yields the desired $M_{\rm env}$.  Thus, for fixed $M_c$ and $M_{\rm env}$, we can describe the evolution of $P_b$ and $T_b$ during the flash as the envelope's entropy is increased by nuclear burning.

\begin{figure}
	\plotone{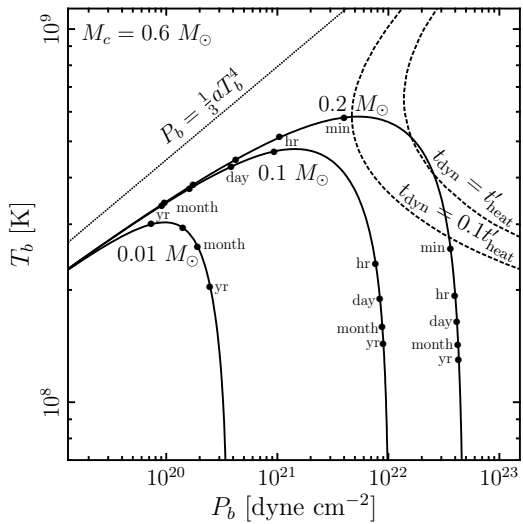}
	\caption{Hydrostatic evolution of $T_b$ and $P_b$ for fully convective $0.01 $, $0.1$, and $0.2 \ M_\odot$ pure helium envelopes on a $0.6 \ M_\odot$ WD with $T_c=10^7$ K (\emph{solid lines}).  The base temperatures at ignition are $\sim 10^8$ K.  Conditions for which the global heating timescale, $t_{\rm heat}$, equals various timescales up to 1 minute are labelled.  As the envelopes evolve to higher entropy configurations, $T_b$ rises at nearly constant pressure until the scale height becomes $\simeq 20\%$ of the WD radius, at which point the envelope expands and $P_b$ and $T_b$ decrease.  The radiation pressure limit, $P_b = a T_b^4/3$ (\emph{dotted line}), provides an upper bound to $T_b$.  Also shown are conditions for which $t_{\rm dyn} = t_{\rm heat}' $ and $0.1t_{\rm heat}' $ (\emph{dashed lines}); see \S \ref{sec:dyn}.}
	\label{fig:basetraj_06}
\end{figure}

\begin{figure}
	\plotone{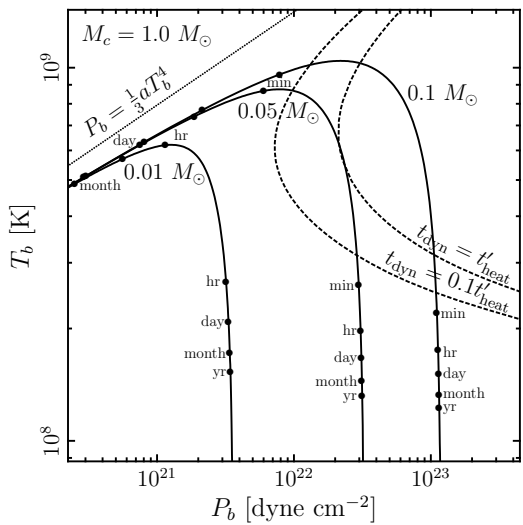}
	\caption{Same as Fig. \ref{fig:basetraj_06}, but for $0.01 $, $0.05$, and $0.1 \ M_\odot$ envelopes on a $1.0 \ M_\odot$ WD.}
	\label{fig:basetraj_10}
\end{figure}

Figure \ref{fig:basetraj_06} shows the evolution of the base temperature and pressure as the entropy increases for $0.01$, $0.1 $, and $0.2 \ M_\odot$ envelopes on a WD with $M_c = 0.6 \ M_\odot$ (\emph{solid lines}).  Figure \ref{fig:basetraj_10} shows the same, but for $0.01$, $0.05$, and $0.1 \ M_\odot$ envelopes on a $1.0 \ M_\odot$ WD.  The envelopes are composed of pure helium for simplicity because the triple-$\alpha$ reaction is the dominant entropy source for displaying the envelope evolution and dynamical burning conditions that are shown in Figures \ref{fig:basetraj_06} and \ref{fig:basetraj_10}; see \S \ref{sec:comp} for details on the effects of envelope composition.

The hydrostatic growth of the envelope is governed by a global heating timescale $t_{\rm heat} \equiv \int ( c_P T dm ) \ / \int ( \epsilon dm )$, where $c_P$ is the specific heat, and the energy generation rate is $\epsilon$.  This timescale is not to be confused with the local heating timescale $t_{\rm heat}' \equiv c_P T_b / \epsilon$; see \S \ref{sec:dyn}.  The integrals are performed over the entire convective envelope because even though the majority of nuclear burning occurs at the base of the envelope, the entropy release must heat the whole convective zone \citep{wbs06,pc08}.  Conditions for which $t_{\rm heat}$ equals a year, a month, a day, an hour, and a minute are labelled in Figures \ref{fig:basetraj_06} and \ref{fig:basetraj_10}, if the envelope evolves quickly enough.  Note that $t_{\rm heat}$ first decreases with increased entropy but then increases after radial expansion becomes important and $T_b$ reaches its peak.

Convection is initiated when $T_b \simeq 10^8$ K.  As He-burning increases the entropy of the envelope, $T_b$ rises at nearly constant pressure because the envelope is initially geometrically thin; the ratio of pressure scale height, $h$, to core radius, assuming pressure due to non-relativistic electron degeneracy, is
\be
	\frac{h}{R_b} &=& 0.07 \lp \frac{P_b}{10^{22} \cgsp } \rp^{2/5} \nonumber \\
	&& \times \lp \frac{M_\odot}{M_c} \rp \lp \frac{R_b}{5\E{8} {\rm \ cm}} \rp ,
\ee
where $h=P_b / \rho_b g$ and $g=GM_c/R_b^2$ is the gravitational acceleration at the envelope base.  Thus, $P_b$ remains nearly constant during the initial phase of the flash and is approximately equal to $ GM_c M_{\rm env}/4 \pi R_b^4$.

\begin{figure}
	\plotone{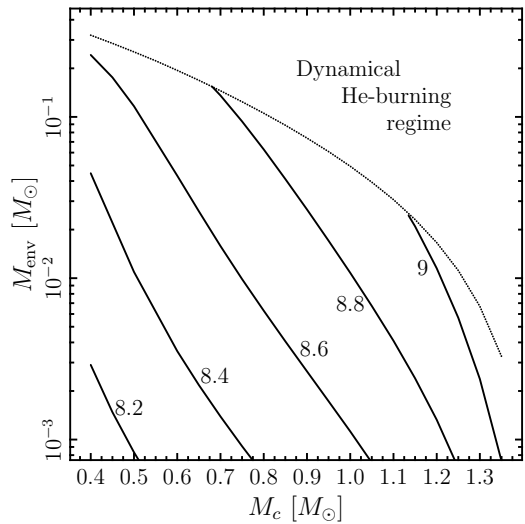}
	\caption{Maximum $T_b$ for convective helium envelopes.  Contours of constant maximum $T_b$ are labelled by their logarithmic value. The minimum $M_{\rm env}$ that achieves dynamical burning (as defined by reaching $t_{\rm dyn} = t_{\rm heat}'$; see \S \ref{sec:dyn}) is shown by the dotted line.  Envelopes larger than this minimum $M_{\rm env}$ become dynamical before reaching their maximum $T_b$.  Lowering the requirement to $t_{\rm dyn} = 0.1 t_{\rm heat}'$ reduces the minimum $M_{\rm env}$ by a factor of $\sim 2$, as shown in Fig. \ref{fig:menvmcore}.}
	\label{fig:Tmax_He}
\end{figure}

The ideal gas pressure and radiation pressure become important as the entropy and temperature increase, and the envelope becomes geometrically thick, resulting in a decrease of $P_b$.  Further entropy increase when the shell is thick results in a decrease of $T_b$, with the maximum temperature achieved when $h \simeq 0.2 R_b$.  This maximum $T_b$ only depends on $M_c$ and $M_{\rm env}$ for a given envelope composition.  Figure \ref{fig:Tmax_He} shows contours of maximum $T_b$ for pure helium envelopes as a function of the WD mass.  The contours are truncated at the $M_{\rm env}$'s above which the burning becomes dynamical before the maximum temperature is reached (\emph{dotted line}); see \S \ref{sec:dyn}.  

Because of the ultracompact nature of AM CVn binaries, we must consider the possibility of binary interactions during the evolution of the helium flash.  At orbital periods $>3 $ min, as expected during the helium flash regime when $\dot{M} = 3\E{-8} - 2\E{-6} \smpy $ \citep{nel01,bild06}, the binary has a separation $> 4\E{9}$ cm, wide enough that the radius of the burning WD stays within its Roche radius for envelopes that achieve dynamical burning.  However, the evolution of non-dynamical envelopes after the peak temperature must account for the impact of Roche-lobe filling and a possible common envelope phase.  We leave these details for future work.


\subsection{Dynamical burning}
\label{sec:dyn}

Our calculations have assumed that the convective motion is very subsonic (i.e., the convection is efficient), so that the envelope is in hydrostatic equilibrium.  However, if the base of the envelope heats up more rapidly than it can adjust to maintain hydrostatic equilibrium, this assumption is contradicted.  To quantify this condition, we compare the local heating timescale at the base, $t_{\rm heat}' \equiv c_P T_b / \epsilon_b $, to the dynamical timescale, $t_{\rm dyn} \equiv h / c_s $, where the pressure scale height, the sound speed, $c_s \equiv \sqrt{ \left. \partial P / \partial \rho \right|_{\rm ad} } $, the specific heat, and the energy generation rate are evaluated at the envelope base.  The local timescale, $t_{\rm heat}'$, as opposed to the global timescale, $t_{\rm heat}$, is the appropriate choice because we wish to know when the base conditions decouple from the rest of the envelope.  The majority of nuclear burning occurs near the base of the envelope, but the thermal energy content is spread out in the envelope, so $t_{\rm heat} ' $ is shorter than $ t_{\rm heat} $ by a factor of $\simeq 10$.

When the local heating timescale becomes shorter than the dynamical timescale and the envelope cannot adjust itself to maintain hydrostatic equilibrium, an overpressure develops, which may lead to a detonation.  As \cite{bswn07} explain, such a detonation at the relatively low ignition pressures achieved in AM CVn systems produces radioactive isotopes such as $^{48}$Cr, $^{52}$Fe, and $^{56}$Ni.  The combination of their short half-lives and the small ejecta masses yields explosions with brightnesses comparable to sub-luminous Type Ia SNe, but with more rapid decays of $\simeq 5-10$ days.

Figures \ref{fig:basetraj_06} and \ref{fig:basetraj_10} show conditions for which $ t_{\rm dyn} = t_{\rm heat}' $ for a $0.6$ and a $1.0 \ M_\odot$ WD, respectively (\emph{dashed lines}).  We also show conditions for which $t_{\rm dyn} = 0.1 t_{\rm heat}'$ in order to account for uncertainty in the exact requirements for the initiation of dynamical burning.  Since dynamical burning is achieved before an envelope reaches its peak $T_b$, the contours in Figure \ref{fig:Tmax_He} are truncated for envelopes that reach $t_{\rm dyn} = t_{\rm heat}'$.

\begin{figure}
	\plotone{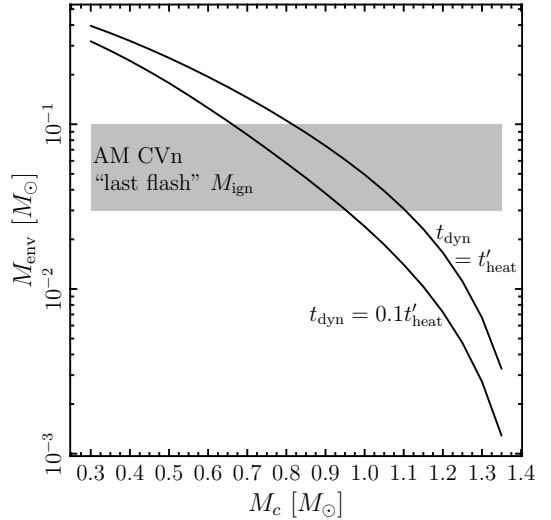}
	\caption{Minimum $M_{\rm env}$'s that reach dynamical burning conditions vs. $M_c$ (\emph{solid lines}).  To account for uncertainties, masses that achieve $t_{\rm dyn} = t_{\rm heat}'$ and $t_{\rm dyn} = 0.1t_{\rm heat}'$ are shown.  The shaded region shows $M_{\rm env} = 0.03 - 0.1 \ M_\odot$, which is an estimate of the ``last-flash'' $M_{\rm ign}$ expected for AM CVn binaries \citep{bswn07}.}
	\label{fig:menvmcore}
\end{figure}

It is clear that large envelopes will burn dynamically, while smaller ones will not, and that there is a minimum envelope mass for which dynamical burning conditions are reached.  Typical timescales at the onset of dynamical burning for these minimum mass envelopes range from $t_{\rm dyn} = t_{\rm heat}' = 0.1-0.5 $ s.  Figure \ref{fig:menvmcore} shows these minimum dynamically-burning $M_{\rm env}$'s vs. WD mass for both $t_{\rm dyn} = t_{\rm heat}'$ and $0.1 t_{\rm heat}'$ (\emph{solid lines}).  As discussed in \S \ref{sec:amcvn_mign}, the values of the ``last-flash'' $M_{\rm ign}$'s for AM CVn systems remain somewhat uncertain due to the as-yet uncalculated effects of binary evolution and core heating.  Thus, we show a range of last-flash $M_{\rm ign} = 0.03 - 0.1 \ M_\odot $ (\emph{shaded region}), which indicates that accretor masses $ \gtrsim 0.8 \ M_\odot$ will undergo dynamical burning in AM CVn systems, giving rise to .Ia SNe.  Large WD masses also likely demand O/Ne WDs, which would obviate any concern of a carbon detonation deep in the accreting WD triggered via shock focusing \citep{liv90,fhr07}.

\subsection{Helium core flash on the red giant branch}
\label{sec:RGB}

The simplicity of our calculation can be applied to any situation involving a convective He-burning shell on an isothermal core, with the caveat that the outer boundary condition at the edge of the convective shell may be different for different circumstances.  In particular, we now briefly examine the helium core flash at the tip of the red giant branch (RGB), which produces convective helium burning shells on a partially degenerate helium core.

As a $<2M_\odot$ star evolves up the RGB, a temperature inversion develops in the growing degenerate helium core \citep{thom67,sg78}.  When the helium core mass reaches $0.48 \ M_\odot$ (see \citealt{sf05} for recent calculations), the temperature peaks at a mass coordinate of $0.3 \ M_\odot$ \citep{fih90}, yielding a convective He-burning shell with $M_c = 0.3 \ M_\odot$ and $M_{\rm env} = 0.18 \ M_\odot$ (e.g., \citealt{sg78,ir84}).  This is the first of $\simeq 10$ core flashes \citep{ir84}, with the mass coordinate of the convective envelope base moving inwards with each flash.  In $\simeq 10^6$ years, the degeneracy has been removed from the helium core, and the star undergoes stable helium core burning on the horizontal branch \citep{sw05}.

\begin{figure}
	\plotone{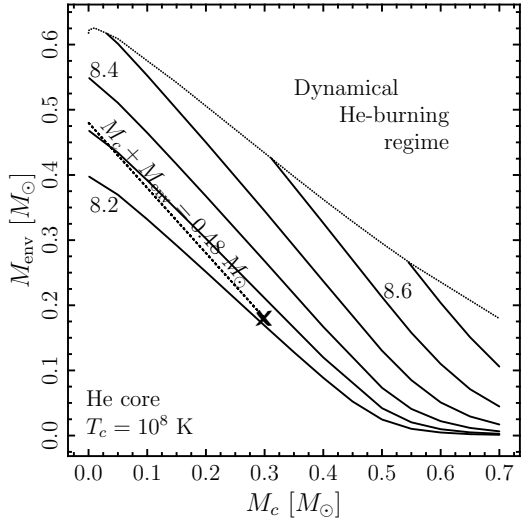}
	\caption{Same as Fig. \ref{fig:Tmax_He}, but for helium envelopes on helium cores with $T_c = 10^8$ K as appropriate for helium core flashes at the tip of the red giant branch, and with peak temperature contours ranging from $10^{8.2} - 10^{8.7}$.  The dashed line shows a constant total mass of $M_c+M_{\rm env} = 0.48 \ M_\odot$, and the cross shows the first core flash, which results in a $0.18 \ M_\odot$ convective envelope on a $0.3 \ M_\odot$ helium core.}
	\label{fig:Tmax_He_on_He}
\end{figure}

Figure \ref{fig:Tmax_He_on_He} shows contours of peak temperature as in Figure \ref{fig:Tmax_He}, but for helium envelopes on a hot helium core with $T_c = 10^8$ K as appropriate for these helium core flashes.  While the exact core composition and $T_c$ do not matter for our other calculations, they are important for these low mass calculations with $M_c < 0.4 \ M_\odot$ because the small hot core is only partially degenerate. The dashed line shows values of $M_c$ and $M_{\rm env}$ that give a constant helium mass of $M_c+M_{\rm env} = 0.48 \ M_\odot$, the mass of the helium region at the tip of the RGB, and the cross shows the core and envelope mass of the first helium core flash.  Our calculation does not take into account the relatively high entropy boundary condition at the top of the helium convective zone, which results from the hydrogen-burning shell above.  However, this boundary condition has very little effect on the conditions at the base of the helium layer; see, e.g., Figure 8 of \cite{thom67}.

Figure \ref{fig:Tmax_He_on_He} shows clearly that the peak temperature of all of the flashes is $\simeq 2\E{8}$ K.  Moreover, we see that these flashes never become dynamical, since the combination of $M_c$ and $M_{\rm env}$ yielding a total mass of $0.48 \ M_\odot$ is always in the hydrostatically-burning regime \citep{fih90}. Thus, helium core flashes should be accurately calculable by presuming hydrostatic balance and efficient convection.  In fact, summing $M_c$ and the minimum dynamically-burning $M_{\rm env}$ in Figure \ref{fig:Tmax_He_on_He} shows that no pure helium star $ \lesssim 0.6 \ M_\odot$ can ever achieve dynamical burning.  This calculation clarifies why modern multi-dimensional hydrodynamic calculations \citep{dle06,mocak08a,mocak08b} nearly always find that the flash leads to convective velocities close to that expected from mixing length and temperature gradients close to the adiabatic gradient.


\section{Envelope composition}
\label{sec:comp}

The previous calculation assumed that the convective envelope consists of pure helium, neglecting the presence of other elements that have been accreted from the donor, produced by nucleosynthesis, or dredged up from the underlying WD core.  These trace elements do not greatly affect the $P_b - T_b$-trajectories of the envelopes because those only depend on the equation of state, which is always dominated by helium.  The composition also does not strongly affect the heating timescales, as the trace elements are depleted when burning conditions are reached.

However, the propagation of detonations may be affected by the presence of $^{14}$N, $^{16}$O, $^{18}$F, $^{20}$Ne, and $^{24}$Mg, which lead to nuclear reactions that are more volatile than the triple-$\alpha$ process at post-shock temperatures $> 10^9$ K.  These isotopes can accelerate the burning and shorten the Zel'dovich-von Neumann-D\"{o}ring (ZND) detonation length.  All previous work has only been for pure helium composition \citep{kho88,kho89,tn00} and thus neglects these effects.  Furthermore, the occurrence of weak interactions changes the neutron-to-proton fraction, impacting the nucleosynthetic outcome of the detonation.  In this section, we explore the range of abundances expected at the onset of dynamical burning in order to motivate future hydrodynamical calculations.


\subsection{Accreted composition}
\label{sec:comp_acc}

The material that is accreted from the degenerate helium donor in an AM CVn system is nearly pure helium.  Assuming the donor has undergone CNO-burning of its hydrogen, the majority of the accreted CNO isotopes will exist as $^{14}$N because the $^{14}$N proton-capture is the slowest step of the CNO cycle for typical CNO-burning conditions.  Thus, we expect the accreted envelope to have a $^{14}$N mass fraction of  $X_{\rm 14N} \simeq 9\E{-3} $ (the sum of solar CNO mass fractions from \citealt{lod03}), a helium mass fraction of $Y=0.99$, and a negligible amount of heavier elements.

Because the triple-$\alpha$ process is active at lower temperatures than $\alpha$-captures on $^{14}$N, the mass fraction of $^{14}$N will not change prior to the onset of thermally-unstable triple-$\alpha$ burning and convection unless conditions are reached for $^{14}$N electron-captures, which are the beginning of the NCO chain: $^{14}$N$(e^-,\nu)^{14}$C$(\alpha,\gamma)^{18}$O$(\alpha,\gamma)^{22}$Ne.  However, the threshold density required for the $e^-$-captures to outpace the reverse $\beta$-decays is $1.2 \E{6} \cgsd $ for a He-dominated plasma \citep{hashi86,pct01}, which corresponds to an electron degeneracy pressure of $4.0 \E{22} \cgsp $.  For $0.65 \ M_\odot < M_c < 1.3 \ M_\odot$, the minimum $M_{\rm env}$'s that achieve $t_{\rm dyn} = t_{\rm heat}'$ begin their convective evolution with $P_b$ lower than this threshold pressure, and thus $X_{\rm 14N}$ will remain at $10^{-2}$.  For envelopes with $P_b > 4.0 \E{22} \cgsp $ at the onset of convection, it is possible that some of the accreted $^{14}$N will capture electrons.  However, the timescales associated with the $^{14}$N electron-capture are $ \gtrsim 10^7$ yr, and thus it is unclear if there is enough time for the conversion of $^{14}$N to $^{14}$C in the $10^6$ yr it takes to accrete $0.1 \ M_\odot$ at $10^{-7} \smpy$ even if the threshold density is reached.  Furthermore, if electron-captures do occur, they will occur only at the bottom of the accreted envelope, where the density is highest.  A proper calculation of the $^{14}$N and $^{14}$C abundances awaits a time-dependent simulation and thus we explore the fate of $^{14}$C in \S \ref{sec:14fate}, but it is likely that the majority of accreted $^{14}$N nuclei will still be $^{14}$N at the onset of convection in an AM CVn system.


\subsection{Composition changes due to core dredge-up}
\label{sec:comp_dredge}

Analytic, numeric, and observational work on classical novae (e.g., \citealt{starr72,pk84,tl86,gehrz98}) show that core material is often entrained in the convective envelope.  The amount of core material can be as high as 30\% by mass in classical nova ejecta, and such a dredge-up would have a large effect on the abundances of $^{12}$C, $^{16}$O, or $^{20}$Ne in the convective helium envelope, depending on the composition of the C/O or O/Ne core material directly underneath the accreted envelope.

However, the exact process by which the classical nova envelope becomes enriched in core material is uncertain (see \citealt{sb09} for a summary), and it is not clear that the same mechanisms also function for helium flashes.  For example, one possible classical nova core-envelope mixing process involves the diffusion of hydrogen into $^{12}$C-rich core material \citep{pk84}.  Convection is then initiated below the bulk of the accreted layer in the diffusing hydrogen tail via $p+^{12}$C reactions.  However, in the case of helium accretion, the diffusion of helium into $^{12}$C-, $^{16}$O-, or $^{20}$Ne-rich material below the layer will not trigger a flash because the reaction rates for $\alpha$-captures on these elements at $\sim 10^8$ K are much slower than the triple-$\alpha$ reaction rate.  Convection is thus almost certainly initiated at or above the base of the accreted layer via triple-$\alpha$ or NCO reactions.

Core-envelope mixing in classical novae is also hypothesized to occur via convective overshoot \citep{woos86} or shear mixing due to the convective motion \citep{glt97,glt07,ros01}.  Such mixing mechanisms have not been studied in the context of convective helium shells, so it is unclear whether or not they can significantly alter the envelope composition.  Since core-envelope mixing is not decisively ruled out, we allow for the possibility of core-envelope mixing in \S \ref{sec:12fate} by considering the fate of $^{12}$C, $^{16}$O, and $^{20}$Ne in the convective envelope.


\subsection{Compositional evolution due to nucleosynthesis}

We now calculate the evolution of $^{14}$N, $^{14}$C, $^{12}$C, $^{16}$O, and $^{20}$Ne due to nuclear burning.  While our methodology prohibits the time-dependent calculation of an isotope's abundance, we can determine its absence or presence by comparing $t_{\rm heat}$ to its depletion timescale, $t_i \equiv \left| d \ln X_i / dt\right|^{-1}$.  The depletion timescale in a uniformly-mixed convective envelope with respect to $\alpha$-captures is
\be
	t_i = \frac{M_{\rm env}}{ \int n_4 \langle \sigma v \rangle_i \ dm}
	\label{eq:ti}
\ee
where the integral is performed over the convective envelope, $n_4$ is the number density of $^4$He, and $ n_4 \langle \sigma v \rangle_i$ is isotope $i$'s $\alpha$-capture rate.  When an isotope is consumed more rapidly but is produced more slowly than the envelope evolves, its abundance drops exponentially.  For example, if the lifetime of $^{16}$O with respect to $\alpha$-captures becomes shorter than $t_{\rm heat}$ but the lifetime of $^{12}$C remains longer than $t_{\rm heat}$, any $^{16}$O present will be converted to $^{20}$Ne (and then possibly to an isotope higher in the $\alpha$-chain) without being replenished by the conversion of $^{12}$C to $^{16}$O.


\subsubsection{The fate of $\alpha$-chain elements}
\label{sec:12fate}

We begin by calculating the lower limit to the $^{12}$C mass fraction, $X_{12}$, set by He-burning.  Since the vast majority of heat is generated by the conversion of $^4$He to $^{12}$C, we can estimate the $^{12}$C mass fraction, $X_{12}$, produced through nucleosynthesis at a given evolutionary stage by calculating the difference in total energy between that stage and that of the envelope at the onset of convection.  This will be a lower limit to the actual $^{12}$C abundance at that time because core-envelope mixing may introduce additional carbon from the core.  Furthermore, we neglect the kinetic energy of convective motion in this estimate, so the actual difference in energies and thus $X_{12}$ may be much larger.

We must take care to include the energy contained in the core as well as the envelope because the significant radial expansion of the envelope during the course of the convective evolution means work is done on the envelope by the core.  The total energy in the WD, neglecting the kinetic energy of convection, is
\be
	E_{\rm total} = \int_0^{M_c+M_{\rm env}} \lp u - \frac{Gm}{r} \rp dm ,
\ee
where $u$ is the specific internal energy.  The minimum $M_{\rm env}$ that achieves dynamical burning for a $0.6 \ M_\odot$ WD is $0.19 \ M_\odot$.  To evolve from an initial state of $T_b=10^8$ K to the stage where $t_{\rm dyn} = t_{\rm heat}'$ requires $\Delta E_{\rm total} \simeq 8\E{48}$ erg, which is made available by burning $7\E{-3} \ M_\odot$ of $^4$He to $^{12}$C, giving a final carbon mass fraction $X_{12} \simeq 0.04 $.  For a $0.05 \ M_\odot$ envelope on a $1.0 \ M_\odot$ WD, a similar analysis yields $X_{12} \simeq 0.05$ at the onset of dynamical burning.  For larger envelopes than these, the energy required to achieve dynamical burning is smaller (because $T_b$ at the onset of dynamical burning is lower; see Figs. \ref{fig:basetraj_06} and \ref{fig:basetraj_10}) and $M_{\rm env}$ is larger, yielding $X_{12} = 10^{-3} - 10^{-2} $ at the onset of dynamical burning.  However, as previously mentioned, these values are a lower limit due to our neglect of the material dredged up from the core and the energy in convective motions.

\begin{figure}
	\plotone{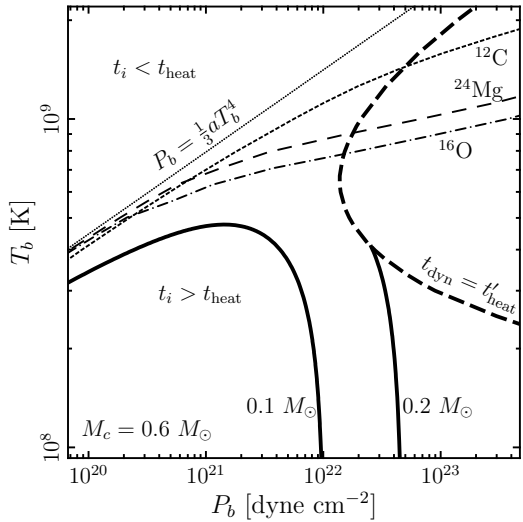}
	\caption{Base temperatures and pressures for which isotope lifetimes, $t_i$, are equal to $t_{\rm heat}$.  The isotopic lifetimes for $^{12}$C (\emph{short-dashed line}), $^{16}$O (\emph{dashed-dotted line}), and $^{24}$Mg (\emph{long-dashed line}) are calculated with respect to $\alpha$-captures.  The lifetimes are longer than $t_{\rm heat}$ for temperatures and pressures below the depletion lines.  Also shown is the radiation pressure limit (\emph{dotted line}) and base trajectories for $0.1 $ and $0.2 \ M_\odot$ pure helium envelopes on a $0.6 \ M_\odot$ WD (\emph{thick solid line}), the latter of which is truncated at the dynamical burning line where $t_{\rm dyn} = t_{\rm heat}'$ (\emph{thick dashed line}).}
	\label{fig:12evol}
\end{figure}

We now consider the fate of trace $\alpha$-chain elements heavier than helium (i.e., $^{12}$C, $^{16}$O, $^{20}$Ne, and $^{24}$Mg).  The lines in Figure \ref{fig:12evol} show the base temperatures and pressures for which $t_i$ of equation \ref{eq:ti} is equal to $ t_{\rm heat}$.  The global isotope depletion and envelope heating timescales are functions of $M_c$, but due to the strong temperature-dependence of nuclear burning, the isotopic depletion lines are nearly independent of $M_c$; for simplicity, we only show lines calculated for $M_c = 0.6 \ M_\odot$.  The unscreened $\alpha$-capture lifetimes for $^{16}$O (\emph{dashed-dotted line}) and $^{24}$Mg (\emph{long-dashed line}) are calculated with the MESA code, while the lifetime of $^{12}$C (\emph{short-dashed line}) is calculated from \cite{kunz02}.  The depletion line for $^{20}$Ne is not shown but is nearly identical to that of $^{16}$O.  The radiation-pressure limit is represented by a dotted line.  Also shown are the trajectories of $0.1$ and $0.2 \ M_\odot$ pure helium envelopes on a $0.6 \ M_\odot$ WD (\emph{thick solid line}).  The trajectory of the $0.2 \ M_\odot$ envelope has been truncated at the dynamical burning line where $t_{\rm dyn} = t_{\rm heat}'$ (\emph{thick dashed line}).  As with the depletion lines, the conditions for which $t_{\rm dyn} = t_{\rm heat}'$ are a function of $M_c$ but depend only slightly on it because nuclear-burning is strongly temperature-sensitive.

Isotope lifetimes are longer than $ t_{\rm heat}$ in the regions below the depletion lines, so an isotope's abundance does not decrease due to $\alpha$-captures significantly in this area of parameter space, although it may increase if a different isotope is converted into it.  In particular, the $^{12}$C mass fraction does not decrease from $\alpha$-captures because the hydrostatic evolution of all envelopes remains below the $^{12}$C depletion line, but $X_{12}$ does increase due to triple-$\alpha$ reactions.

Figure \ref{fig:12evol} shows that depletion of $\alpha$-chain isotopes occurs in a region of parameter space inaccessible to hydrostatically-burning convective envelopes, and thus they are still unburned when the envelope becomes dynamical.  If a shock does develop and the post-shock temperature is $ \gtrsim 10^9$ K, the unburned $\alpha$-chain elements, and especially $^{16}$O, $^{20}$Ne, and $^{24}$Mg, capture $\alpha$-particles faster than the triple-$\alpha$ process operates.  Thus, hydrodynamic simulations of helium detonations must include unburned $\alpha$-chain elements resulting from dredge-up and nucleosynthesis, as they can reduce the ZND length as compared to pure helium detonations.


\subsubsection{The fate of $^{14}${\rm N} and $^{14}${\rm C} in envelopes that become dynamical}
\label{sec:14fate}

We now calculate the evolution of the accreted $^{14}$N, a small amount of which will be converted to $^{14}$C if $\rho_b>10^6 \cgsd$ and the accretion timescale $\gtrsim 10^7 $ yr.  Due to the $\beta$-decays in their reaction chains, these isotopes determine the neutron-to-proton ratio at the onset of dynamical burning, which will affect the isotopic yield of the detonation.  Furthermore, as we will see, they can affect the progress of the helium detonation by providing protons to catalyze other reactions.  These isotopes will also play a role in the nucleosynthetic outcome of envelopes that do not achieve dynamical burning, but we leave these details for future work.

\begin{figure}
	\plotone{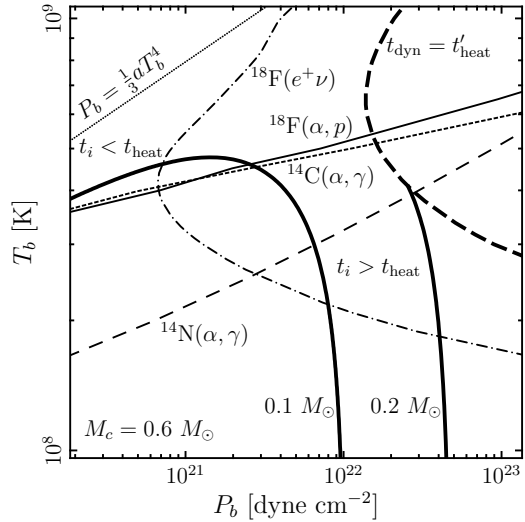}
	\caption{Same as Fig. \ref{fig:12evol}, but with different isotopes: the isotopic lifetimes for $^{14}$N (\emph{long-dashed line}) and $^{14}$C (\emph{short-dashed line}) are calculated with respect to $\alpha$-captures, while the lifetime of $^{18}$F is shown with respect to both $\alpha$-captures (\emph{solid line}) and $\beta$-decays (\emph{dashed-dotted line}).}
	\label{fig:14evol}
\end{figure}

Conditions for which $t_{\rm heat}$ equals the positron-emission lifetime of $^{18}$F (\emph{dashed-dotted line}) and the unscreened $\alpha$-capture lifetimes for $^{14}$N (\emph{long-dashed line}), $^{14}$C (\emph{short-dashed line}, \citealt{gor92}), and $^{18}$F, which yields $^{21}$Ne and a proton (\emph{solid line}; \citealt{kar08}), are shown in Figure \ref{fig:14evol}.  The other lines are as in Figure \ref{fig:12evol}, with the same caveat that while the calculations shown are for $M_c=0.6 \ M_\odot$, they are nearly identical for other core masses due to the strong temperature-sensitivity of nuclear burning.  It is clear that any $^{14}$C that is produced prior to convection will remain as $^{14}$C at the onset of dynamical burning.  Decays back to $^{14}$N have a half-life of $ 5730 $ yr (or even longer due to exclusion principle inhibition at the base of the envelope; \citealt{pb63,hashi86}), which is much longer than the convective burning phase.  Furthermore, $\alpha$-captures are negligible because the $^{14}$C depletion line crosses the dynamical burning line at a pressure $ < 4\E{22} \cgsp$; i.e., $\alpha$-captures on $^{14}$C at the high densities required to create $^{18}$O occur at temperatures $>5.5 \E{8}$ K, higher than necessary for dynamical burning at these densities.

Likewise, $^{14}$N will remain unchanged prior to dynamical burning in nearly all cases.  If conditions are reached for $^{14}$N$(\alpha,\gamma)^{18}$F reactions, the $^{18}$F will not have a chance to $\beta$-decay because its $\beta$-decay half-life of 110 min is much longer than the timescale at the onset of dynamical burning, $t_{\rm dyn} = t_{\rm heat}'$.  Electron-captures onto $^{18}$F during the hydrostatic envelope evolution can also be neglected, as their timescale is much longer than $t_{\rm heat}$ for conditions after $^{14}$N $\alpha$-captures have taken place \citep{bah64,lang80,twcc95}.  Therefore, no weak interactions will occur prior to dynamical burning for nuclei that are $^{14}$N at the onset of convection, even if they are converted to $^{18}$F.

Thus, the neutron-to-proton ratio at the onset of dynamical burning is determined at the beginning of convection: nucleosynthesis during the convective phase will not change this ratio before dynamical burning occurs.  If $^{14}$C exists prior to convection, it will remain $^{14}$C, and $^{14}$N will either remain unburned or capture an $\alpha$-nucleus to become $^{18}$F, but will not undergo any weak interactions.

As with the $\alpha$-chain isotopes, it is important to consider the presence of $^{14}$C and $^{18}$F when calculating the hydrodynamic evolution of the detonation.  At temperatures $>10^9$ K, the $\alpha$-capture chain $^{14}$C$(\alpha,\gamma)^{18}$O$(\alpha,\gamma)^{22}$Ne will take place, although its energy input will likely be negligible due to the small abundance of $^{14}$C.  More significantly, $\alpha$-captures on $^{18}$F lead to the production of protons via $^{18}$F$(\alpha,p)^{21}$Ne (recently examined in the context of AGB evolution in \citealt{kar08}), which act as a catalyst for $\alpha$-chain reactions via $^{12}$C$(p,\gamma)^{13}$N$(\alpha,p)^{16}$O \citep{wbs06}.  This process bypasses the comparatively slow $^{12}$C $\alpha$-capture and will certainly impact the evolution of the detonation.  Furthermore, this reaction will play a role in determining the nucleosynthetic outcome of envelopes that do not become dynamical, the details of which we leave to future work.


\section{Conclusion}
\label{sec:conc}

Motivated by \cite{bswn07}'s result that dynamical helium burning is naturally attained during the evolution of AM CVn binaries, we have studied the progress of convective helium envelopes via hydrostatic calculations under simplifying assumptions (\S \ref{sec:calc}).  We have found the maximum possible temperatures (Fig. \ref{fig:Tmax_He}) as well as the minimum envelope masses that achieve dynamical burning, as defined when the local heating timescale, $t_{\rm heat}'$, equals the sound-crossing time of the envelope, $t_{\rm dyn}$ (Fig. \ref{fig:menvmcore}).  The generality of our calculation allowed us to also examine the helium core flash at the tip of the red giant branch (\S \ref{sec:RGB}), confirming previous results that the flash is a slow enough event that hydrostatic evolution is a safe assumption.

In order to motivate the starting conditions for future hydrodynamic studies, we examined the nucleosynthetic fate of elements in \S \ref{sec:comp}.  We found that $\alpha$-chain elements such as $^{12}$C, $^{16}$O, $^{20}$Ne, and $^{24}$Mg, produced via nucleosynthesis or dredged up from the degenerate core, remain unburned during the hydrostatic envelope evolution (Fig. \ref{fig:12evol}) and will be present if a detonation commences, affecting the ZND length.

For large envelope and core masses, we have found that some $^{14}$N can capture electrons during the accretion phase to form $^{14}$C, but this is likely a small effect.  Independently of envelope and core masses, we have found that the neutron-to-proton ratio at the onset of dynamical burning, which will determine the detonation's isotopic distribution, is the same as at the onset of convection because there is no opportunity for weak interactions during the convective evolution to dynamical burning; the timescale for the obvious candidate (positron-emission by $^{18}$F) is so long that envelopes that achieve dynamical burning do so before $^{18}$F can $\beta$-decay (Fig. \ref{fig:14evol}).  The existence of $^{18}$F is very important for determining the ZND length because of the production of protons via $^{18}$F$(\alpha,p)^{21}$Ne, which will bypass the slow step of the $\alpha$-chain by catalyzing the conversion of $^{12}$C to $^{16}$O.


\acknowledgments
 
We thank G. Nelemans and N. Weinberg for helpful discussions and B. Paxton for assistance with MESA.  This work was supported by the National Science Foundation under grants PHY 05-51164 and AST 07-07633.


\appendix

\section{Chemical diffusion between the WD core and envelope}\
\label{sec:diff}

In this Appendix, we justify our assumption of a discontinuous composition profile at the WD core-envelope interface. Due to electron degeneracy, there is a negligible differentiating force on helium in a degenerate C/O or O/Ne plasma. Hence, the accreted helium simply diffuses into the WD core at the rate given by a random walk and penetrates a distance $ r_{\rm in} \simeq \sqrt{ D t_{\rm acc} }$ during the time it takes to accumulate an ignition mass, $t_{\rm acc} = M_{\rm ign} / \dot{M}$, where $D$ is the diffusion coefficient.  For a carbon plasma, the Coulomb coupling parameter of the one component plasma (OCP), which measures the strength of ion-ion interactions, is $\Gamma_{12,12} = 1.7 \rho_5^{1/3} / T_8$, where $\rho_5=\rho/10^5 \cgsd $ and $T_8 = T/10^8 $ K.  Thus, the core-envelope interface is in an intermediate regime between weak and strong ion coupling, for which an exact value for $D$ at these temperatures and densities has not yet been calculated for the case of helium diffusion in a carbon plasma \citep{ppfm86}.  The fitting formula of \cite{im85} yields $D = 0.03 {\ \rm cm^2 \ s^{-1} \ } T_8^{5/2} / \rho_5 $, neglecting the temperature- and density-dependence of the Coulomb logarithm.  This estimate yields
\be
	\frac{r_{\rm in}}{h} \sim 0.01 \frac{g_8 T_8^{5/4} }{\rho_5^{7/6} }  \lp \frac{t_{\rm acc}}{10^6 {\rm \ yr}} \rp^{1/2},
\ee
where the pressure scale height near the core-envelope interface is $h=P_b/\rho_b g$, and $g_8$ is the gravitational acceleration, $g$, in units of $10^8$ cm s$^{-2}$. Thus, diffusion is unimportant for $t_{\rm acc}<10^8 $ yr, or $\dot{M}>10^{-9} \smpy$ for $M_{\rm ign} \leq 0.1 \ M_\odot$, as long as the diffusion coefficient is not actually  orders of magnitude higher than given by \cite{im85}'s fitting formula.

An alternate estimate for $D$ can be made by appealing to results for the OCP.  No appropriate calculations are available for trace helium in a C/O OCP, but we could
attempt to follow Bildsten and Hall (2001) by using a Stokes-Einstein-like argument, assigning a finite size to the helium ion. However, \cite{dal06} showed that the expected Stokes-Einstein relation between $D_{\rm OCP}$ and the viscosity of the medium, $\eta$, only begins to hold in the OCP once $\Gamma \gtrsim 40$. Hence, our only real choice for an
alternate estimate is to use an approximate result relevant for $\Gamma \simeq 1-10$, $D_{\rm OCP} \simeq 3\omega_p a^2 \Gamma^{-4/3}$ \citep{hmp75,dal06}, where $\omega_p^2 = 4\pi \rho (e/2m_p)^2$ is the square of the ion plasma frequency, and $a$ is the average ion spacing in the OCP.  This yields $ D_{\rm OCP} = 0.03 {\ \rm cm^2 \ s^{-1} \ } T_8^{4/3} / \rho_5^{31/54} $, and thus nearly the same result that diffusion is insignificant for $t_{\rm acc} < 10^8 $ yr.




\begin{thebibliography}

\bibitem[Alastuey \& Jancovici(1978)]{aj78}
Alastuey, A., \& Jancovici, B. 1978, \apj, 226, 1034

\bibitem[Alcock \& Illarionov(1980)]{ai80}
Alcock, C., \& Illarionov, A. 1980, \apj, 235, 534

\bibitem[Bahcall(1964)]{bah64}
Bahcall, J. N. 1964, \apj, 139, 318

\bibitem[Bildsten et al.(2007)]{bswn07}
Bildsten, L., Shen, K. J., Weinberg, N. N., \& Nelemans, G. 2007, \apj, 662, L95

\bibitem[Bildsten et al.(2006)]{bild06}
Bildsten, L., Townsley, D. M., Deloye, C. J., \& Nelemans, G. 2006, \apj, 640, 466

\bibitem[Chapman \& Cowling(1952)]{cc52}
Chapman, S., \& Cowling, T. G. 1952, Mathematical Theory of Non-Uniform Gases (Cambridge; Cambridge University Press)

\bibitem[Clayton et al.(2007)]{clay07}
Clayton, G. C., Geballe, T. R., Herwig, F., Fryer, C., \& Asplund, M. 2007, \apj, 662, 1220

\bibitem[Daligault(2006)]{dal06}
Daligault, J. 2006, PRL, 95, 065003

\bibitem[Dearborn et al.(2006)]{dle06}
Dearborn, D. S. P., Lattanzio, J. C., \& Eggleton, P. P. 2006, \apj, 639, 405

\bibitem[Eggleton(1983)]{egg83}
Eggleton, P. P. 1983, \apj, 268, 368

\bibitem[Fink et al.(2007)]{fhr07}
Fink, M., Hillebrandt, W., \& R\"{o}pke, F. K. 2007, \aap, 476, 1133

\bibitem[Foley et al.(2009)]{fol09}
Foley, R. J., et al. 2009, astro-ph/0902.2794

\bibitem[Fujimoto(1982)]{fuji82}
Fujimoto, M. Y. 1982, \apj, 257, 752

\bibitem[Fujimoto et al.(1990)]{fih90}
Fujimoto, M. Y., Iben, I., Jr., \& Hollowell, D. 1990, \apj, 349, 580

\bibitem[Fujimoto \& Sugimoto(1982)]{fs82}
Fujimoto, M. Y., \& Sugimoto, D. 1982, \apj, 257, 291

\bibitem[Gehrz et al.(1998)]{gehrz98} 
Gehrz, R. D., Truran, J. W., Williams, R. E., \&  Starrfield, S. 1998, PASP, 110, 3 

\bibitem[Glasner et al.(1997)]{glt97}
Glasner, S. A., Livne, E., \& Truran, J. W. 1997, \apj, 475, 754

\bibitem[Glasner et al.(2007)]{glt07}
Glasner, S. A., Livne, E., \& Truran, J. W. 2007, \apj, 665, 1321

\bibitem[Goriely et al.(2002)]{gor02}
Goriely, S., Jos\'{e}, J., Hernanz, M., Rayet, M., \& Arnould, M. 2002, \aap, 383, L27

\bibitem[G\"{o}rres et al.(1992)]{gor92}
G\"{o}rres, J., et al. 1992, Nucl. Phys. A, 548, 414

\bibitem[Graboske et al.(1973)]{grab73}
Graboske, H. C., Dewitt, H. E., Grossman, A. S., \& Cooper, M. S. 1973, \apj, 181, 457

\bibitem[Hansen et al.(1975)]{hmp75}
Hansen, J. P., McDonald, I. R., \& Pollock, E. L. 1975, Phys. Rev. A, 11, 1025

\bibitem[Harris et al.(1983)]{har83}
Harris, M. J., Fowler, W. A., Caughlan, G. R., \& Zimmerman, B. A. 1983, \araa, 21, 165

\bibitem[Hashimoto et al.(1986)]{hashi86}
Hashimoto, M., Nomoto, K., Arai, K., \& Kaminisi, K. 1986, \apj, 307, 687.

\bibitem[Iben(1975)]{iben75}
Iben, I., Jr. 1975, \apj, 196, 525

\bibitem[Iben \& Renzini(1984)]{ir84}
Iben, I., Jr., \& Renzini, A. 1984, Phys. Rep., 105, 329

\bibitem[Iben \& MacDonald(1985)]{im85}
Iben, I., Jr., \& MacDonald, J. 1985, \apj, 296, 540

\bibitem[Iben \& Tutukov(1989)]{it89}
Iben, I., Jr., \& Tutukov, A. V. 1989, \apj, 342, 430

\bibitem[Iben \& Tutukov(1991)]{it91}
Iben, I., Jr., \& Tutukov, A. V. 1991, \apj, 370, 615

\bibitem[Iglesias \& Rogers(1993)]{ir93}
Iglesias, C. A., \& Rogers, F. J. 1993, \apj, 412, 752

\bibitem[Iglesias \& Rogers(1996)]{ir96}
Iglesias, C. A., \& Rogers, F. J. 1996, \apj, 464, 943

\bibitem[Iijima \& Nakanishi(2008)]{in08}
Iijima, T., \& Nakanishi, H. 2008, \aap, 482, 865

\bibitem[Itoh et al.(1996)]{itoh96}
Itoh, N., Hayashi, H., Nishikawa, A., \& Kohyama, Y. 1996, \apjs, 102, 411

\bibitem[Itoh et al.(1979)]{itoh79}
Itoh, N., Totsuji, H., Ichimaru, S., \& Dewitt, H. E. 1979, \apj, 234, 1079

\bibitem[Karakas et al.(2008)]{kar08}
Karakas, A. I., Lee, H. Y., Lugaro, M., G\"{o}rres, J., \& Wiescher, M. 2008, \apj, 676, 1254

\bibitem[Kato \& Hachisu(2003)]{kh03}
Kato, M., \& Hachisu, I. 2003, \apj, 598, L107

\bibitem[Kato et al.(2008)]{kato08}
Kato, M., Hachisu, I., Kiyota, S., \& Saio, H. 2008, \apj, 684, 1366

\bibitem[Khokhlov(1988)]{kho88}
Khokhlov, A. M. 1988, \apss, 149, 91

\bibitem[Khokhlov(1989)]{kho89}
Khokhlov, A. M. 1989, \mnras, 239, 785

\bibitem[Kunz et al.(2002)]{kunz02}
Kunz, R., et al. 2002, \apj, 567, 643

\bibitem[Lang(1980)]{lang80}
Lang, K. R. 1980, Astrophysical Formulae (Berlin : Springer)

\bibitem[Limongi \& Tornamb\'{e}(1991)]{lt91}
Limongi, M., \& Tornamb\'{e}, A. 1991, \apj, 371, 317

\bibitem[Livne(1990)]{liv90}
Livne, E. 1990, \apj, 354, L53

\bibitem[Livne \& Glasner(1990)]{lg90}
Livne, E. \& Glasner, A. S. 1990, \apj, 361, 244

\bibitem[Livne \& Glasner(1991)]{lg91}
Livne, E. \& Glasner, A. S. 1991, \apj,  370, 272

\bibitem[Lodders(2003)]{lod03}
Lodders, K. 2003, \apj, 591, 1220

\bibitem[Moc\'{a}k et al.(2008a)]{mocak08a}
Moc\'{a}k, M., M\"{u}ller, E., Weiss, A., \& Kifonidis, K. 2008, \aap, 490, 265

\bibitem[Moc\'{a}k et al.(2008b)]{mocak08b}
Moc\'{a}k, M., M\"{u}ller, E., Weiss, A., \& Kifonidis, K. 2008, astro-ph/0811.4083

\bibitem[Nauenberg(1972)]{nau72}
Nauenberg, M. 1972, \apj, 175, 417

\bibitem[Nelemans et al.(2001)]{nel01}
Nelemans, G., Portegies Zwart, S. F., Verbunt, F., \& Yungelson, L. R. 2001, \aap, 368, 939

\bibitem[Nomoto(1982a)]{nom82a}
Nomoto, K. 1982, \apj, 253, 798

\bibitem[Nomoto(1982b)]{nom82b}
Nomoto, K. 1982, \apj, 257, 780

\bibitem[Nomoto \& Sugimoto(1977)]{ns77}
Nomoto, K., \& Sugimoto, D. 1977, \pasj, 29, 765

\bibitem[Paquette et al.(1986)]{ppfm86}
Paquette, C., Pelletier, C., Fontaine, G., \& Michaud, G. 1986, \apjs, 61, 177

\bibitem[Peterson \& Bahcall(1963)]{pb63}
Peterson, V. L., \& Bahcall, J. N. 1963, \apj, 138, 437

\bibitem[Piersanti et al.(1999)]{pier99}
Piersanti, L., Cassisi, S., Iben, I., Jr., \& Tornamb\'{e}, A. 1999, \apj, 521, L59

\bibitem[Piersanti, Cassisi, \& Tornamb\'{e}(2001)]{pct01}
Piersanti, L., Cassisi, S., \& Tornamb\'{e}, A. 2001, \apj, 558, 916

\bibitem[Piro \& Chang(2008)]{pc08}
Piro, A. L., \& Chang, P. 2008, \apj, 678, 1158

\bibitem[Prialnik \& Kovetz(1984)]{pk84}
Prialnik, D., \& Kovetz, A. 1984, \apj, 281, 367

\bibitem[Rogers \& Nayfonov(2002)]{rn02}
Rogers, F. J., \& Nayfonov, A. 2002, \apj, 576, 1064

\bibitem[Rosner et al.(2001)]{ros01}
Rosner, R., Alexakis, A., Young, Y.-N., Truran, J. W., \& Hillebrandt, W. \apj, 562, L177

\bibitem[Saumon et al.(1995)]{scvh95}
Saumon, D., Chabrier, G., \& van Horn, H. M. 1995, \apjs, 99, 713

\bibitem[Serenelli \& Fukugita(2005)]{sf05}
Serenelli, A. M., \& Fukugita, M. 2005, \apj, 632, L33

\bibitem[Serenelli \& Weiss(2005)]{sw05}
Serenelli, A., \& Weiss, A. 2005, \aap, 442, 1041

\bibitem[Shen \& Bildsten(2007)]{sb07}
Shen, K. J., \& Bildsten, L. 2007, \apj, 660, 1444

\bibitem[Shen \& Bildsten(2009)]{sb09}
Shen, K. J., \& Bildsten, L. 2009, \apj, 692, 324

\bibitem[Starrfield et al.(1972)]{starr72}
Starrfield, S., Truran, J. W., Sparks, W. M., \& Kutter, G. S. 1972, \apj, 176, 169

\bibitem[Sweigart \& Gross(1978)]{sg78}
Sweigart, A. V., \& Gross, P. G. 1978, \apjs, 36, 405

\bibitem[Taam(1980a)]{taam80a}
Taam, R. E. 1980, \apj, 237, 142

\bibitem[Taam(1980b)]{taam80b}
Taam, R. E. 1980, \apj, 242, 749

\bibitem[Thomas(1967)]{thom67}
Thomas, H.-C. 1967, Z. Astrophys., 67, 420

\bibitem[Tilley et al.(1995)]{twcc95}
Tilley, D. R., Weller, H. R., Cheves, C. M., \& Chasteler, R. M. 1995, Nucl. Phys. A, 595, 1

\bibitem[Timmes(1999)]{tim99}
Timmes, F. X. 1999, \apjs, 124, 241

\bibitem[Timmes \& Niemeyer(2000)]{tn00}
Timmes, F. X., \& Niemeyer, J. C. 2000, \apj, 537, 993

\bibitem[Timmes \& Swesty(2000)]{ts00}
Timmes, F. X., \& Swesty, F. D. 2000, \apjs, 126, 501

\bibitem[Truran \& Livio(1986)]{tl86}
Truran, J. W., \& Livio, M. 1986, \apj, 308, 721

\bibitem[Tutukov \& Yungelson(1996)]{ty96}
Tutukov, A., \& Yungelson, L. 1996, \mnras, 280, 1035

\bibitem[van Teeseling et al.(1997)]{vt97}
van Teeseling, A., Reinsch, K., Hessman, F. V., \& Beuermann, K. 1997, \aap, 323, L41 

\bibitem[Weinberg et al.(2006)]{wbs06}
Weinberg, N. N., Bildsten, L., \& Schatz, H. 2006, \apj, 639, 1018

\bibitem[Woosley(1986)]{woos86}
Woosley, S. E. 1986, in Nucleosynthesis and Chemical Evolution, ed. B. Hauck, A. Maeder, \& G. Magnet (Sauverny: Geneva Observatory)

\bibitem[Woosley, Taam, \& Weaver(1986)]{wtw86}
Woosley, S. E., Taam, R. E., \& Weaver, T. A. 1986, \apj, 301, 601

\bibitem[Woosley \& Weaver(1994)]{ww94}
Woosley, S. E., \& Weaver, T. A. 1994, \apj, 423, 371

\bibitem[Yungelson(2008)]{yung08}
Yungelson, L. R. 2008, Astronomy Letters, 34, 620

\bibitem[Yoon \& Langer(2004)]{yl04}
Yoon, S.-C., \& Langer, N. 2004, \aap, 419, 645

\end{thebibliography}
\end{document}